\documentclass[aps,letterpaper,amssymb,twocolumn,prl,floatfix]{revtex4}
\usepackage{amssymb}
\usepackage[usenames, dvips]{color}
\usepackage{graphicx}
\begin{document} 

\title{Front Propagation of Spatio-temporal Chaos}
\author{J. W. Kim$^\dagger$, J. Y. Vaishnav$^\dagger$, E. Ott$^\dagger$, 
S. C. Venkataramani$^\ddagger$, and W. Losert$^\dagger$}
\address{$^\dagger$Department of Physics, 
University of Maryland, College Park, Maryland  20742} 
\address{$^\ddagger$Department of Mathematics, 
University of Chicago, Chicago, Illinois 60637} 
\date{\today}

\begin{abstract}
  We study the dynamics of the front separating a spatio-temporally chaotic 
region from a stable steady region using a simple model applicable to 
periodically forced systems. 
In particular, we investigate both the coarsening of the front induced by
the inherent `noise' of the chaotic region, and the long wavelength dynamics
causing the front to develop cusps.
\end{abstract}
\pacs{PACS numbers: }
\vspace{0.5cm}
]
\maketitle 

\section{INTRODUCTION}
  In this paper we study the dynamics of the front separating a 
spatio-temporally chaotic state from a stable steady ordered state. 
Such situations occur in many experimental settings. 
In an experiment on a vertically vibrating granular monolayer of spheres
\cite{Losert1} 
both a state at rest on the plate and a chaotically bouncing state are
stable. When a small perturbation is applied to the stationary state,
the chaotic state is observed to invade the stable state through a propagating 
front. In a Rayleigh-B\'{e}nard convection experiment \cite{Melnikov1}
both straight rolls and spiral
defect chaos are stable under some conditions and it is observed that a 
region of straight rolls is invaded by a region of spiral defect turbulence. 

  For our study we employ a type of model called a Continuum Coupled Map (CCM)
introduced in Ref. \cite{Shankar1}.
Models of this type (Sec. II) are appropriate to periodically forced systems 
(such as 
that in the experiment of the Ref. \cite{Losert1}). In common with other
generic models, like the complex Ginzburg-Landau equation 
or the Swift-Hohenberg equation, 
CCM models are meant to incorporate the minimal basic properties capable of 
reproducing the phenomena of interest. With this in mind we construct our CCM
model to incorporate the essential feature that both a stable steady 
homogeneous attractor and a spatio-temporally chaotic attractor exist.
Using our CCM model, we numerically investigate
two phenomena: (i) the coarsening of the front due to the inherent `noise' 
associated with the spatio-temporal chaos (Sec. III), and 
(ii) cusp formation induced by initial long wavelength perturbations of 
the front location from the flat state (Sec. IV).

  With respect to (i), an important concept used to study various coarsening
processes is scaling. For a large number of systems (e.g., see Ref. 
\cite{Barabasi1}), it is found that the interface width due to roughening, 
$ w(t) $, increases as a power of time, $ w(t) \sim t^\beta $.
The width eventually saturates at a value that increases as a power of
the system size, $ w(L_x) \sim L_x^\alpha $. These scaling properties are 
also observed in our model, and we determine and discuss the scaling exponents
$\alpha$ and $\beta$ that we find.

  With respect to (ii), we argue that on long scale (i.e., long compared to
$w(t)$), our fronts propagate at constant speed in a direction locally 
normal to the interface.
We show that this basic property explains the mechanism of cusp formation and
the evolution of the shape of the fronts observed in our numerical simulations.

\begin{center}
\begin{figure}[b!]
\begin{centering}
\includegraphics[width= 6 cm]{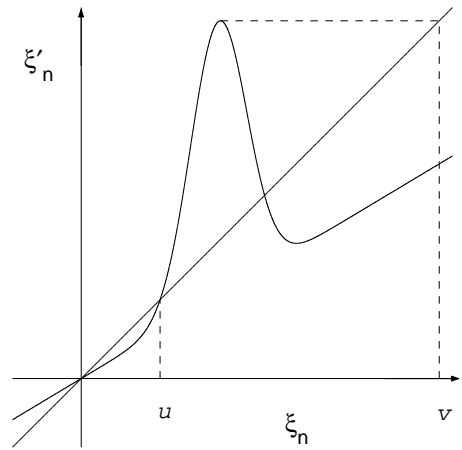}
\end{centering}
\caption{Schematic (not to scale) of the model map.}
\label{fig:map}
\end{figure}
\end{center}

\begin{figure*}[ht!]
\includegraphics[width= 18 cm]{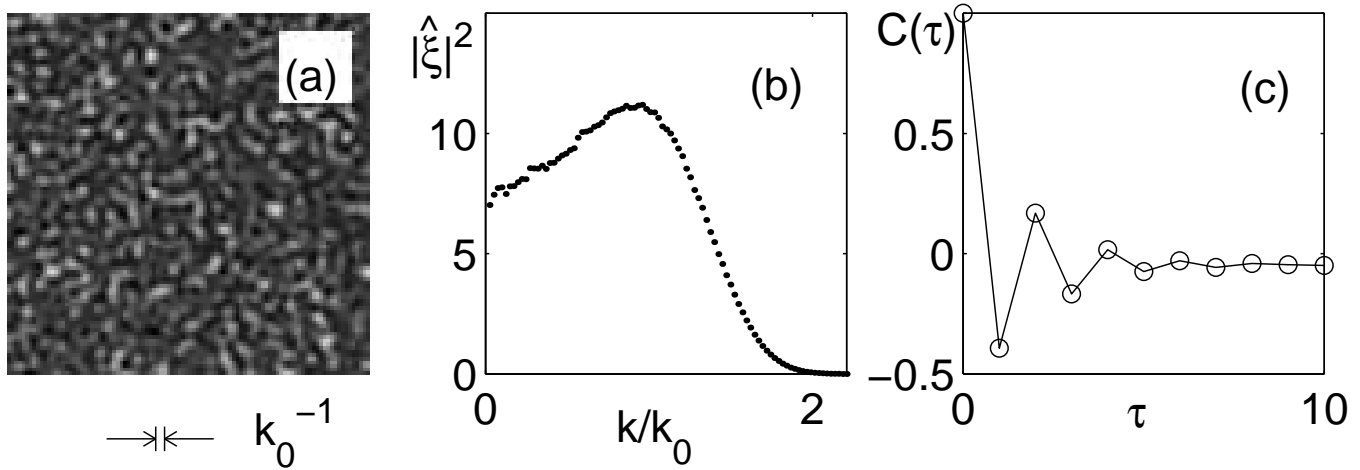}
\caption{Spatio-temporal chaos of CCM model for $ A = 7.0, r = 0.4, $ and $
w = 0.29 $: (a) A snapshot of $ \xi_n (\vec{x})$. Bright regions indicate
large amplitude. (b) Wavenumber power spectrum, averaged over 100 frames. 
$ \hat{\xi} (\vec{k}) $ is the Fourier transformation of $\xi (\vec{x})$.
(c) Time correlation function $C(\tau$), Eq. (\ref{eq:tcor}).}
\label{fig:stchaos1}
\end{figure*}

\section{MODEL} 
  As in reference \cite{Shankar1}, we consider a CCM model which maps a field
$ \xi_n (\vec{x}) $ forward from time $ n $ to time $ n+1 $. With reference to 
a system driven periodically in time (e.g., as in \cite{Losert1}), we may 
think of $ \xi_n(\vec{x}) $, with $n=1,2,\cdots ,$ as being the system state 
stroboscopically sampled once each period. Furthermore, we consider $\vec{x}
=(x,y)$ to be two-dimensional, and, for simplicity, we take $\xi_n$ to be a
scalar field. The CCM model mapping $\xi_n$ to $\xi_{n+1}$ consists of two
steps: The first step is a nonlinear local operation in which a one-dimensional
map $ M $ is applied to $\xi_n(\vec{x})$ at each point in space,
\begin{equation}
\xi'_n(\vec{x})= M(\xi_n(\vec{x})).
\label{eq:map1}
\end{equation}
The second step is a translationally invariant linear operation coupling the 
dynamics at nearby spatial locations. The most general such coupling is 
conveniently expressed in terms of the spatial Fourier transform. If 
$\hat{\xi}_n(\vec{k}) $ is the spatial Fourier transform of
$ \xi_n (\vec{x}) $, then we write $\hat{\xi}_{n+1}(\vec{k})$ as
\begin{equation}
\hat{\xi}_{n+1}(\vec{k}) = \check{f}(\vec{k})\hat{\xi}'_n (\vec{k}),
\label{eq:couple1}
\end{equation}
from which $\xi_{n+1}(\vec{x})$ is obtained by inverse Fourier transforming.

  The model is then specified by the choice of the nonlinear map $M$  and
the linear spatial coupling $\check{f}(\vec{k})$. We make these choices so
as to include the minimum properties that we hypothesize are relevant for 
the investigated phenomena \cite{Description1}.
Since we desire the simultaneous existence of a stable steady state as well as
a spatio-temporally chaotic state, we choose the map $M$ to have a stable
fixed point attractor and a chaotic attractor. A convenient choice having this
property is given by
\begin{equation}
M(\xi) =  r \xi + A {\rm exp}[-(\xi -1)^2/\sigma^2] - A {\rm exp}(-1/\sigma^2).
\label{eq:map2}
\end{equation}
Referring to Fig. \ref{fig:map}, we see that this map has a stable fixed 
point at $\xi = 0$, for $r < 1$. 
Moreover, any initial point in $\xi < u $ is attracted
to this point. We also see from Fig. \ref{fig:map} that the interval
$u<\xi<v$ is mapped into itself. Thus there is (at least) one attractor in this
interval. For the parameter values $A, r$, and $w$ that we investigate
($A=7.0, r=0.4, \sigma=0.29$) there is one attractor in $ u<\xi<v$ 
and it is chaotic.

  Our choice of the spatial coupling $\check{f}(\vec{k})$ is similarly 
motivated by a desire for simplicity. We assume that the coupling is isotropic. 
Thus we can write $\check{f}$ as $\check{f}(\vec{k})=f(k)$, where 
$ k=|\vec{k}|$. Taking $f(k) \geq 0$ we write 
\begin{equation}
f(k)={\rm exp}[\gamma(k)],
\label{eq:fk}
\end{equation}
where $\gamma(k)$ is a wavenumber-dependent growth/damping rate per period. 
Since we want the spatio-temporal
chaos to have a finite spatial correlation scale, such a scale must be
reflected in our choice of $\gamma(k)$. Denoting this scale by $k_0^{-1}$, we
make the following simple choice \cite{Shankar1} for $\gamma(k)$,
\begin{equation}
\gamma (k) = \frac{1}{2} \left(\frac{k}{k_0} \right)^2 \left[ 1- \frac{1}{2} 
\left(\frac{k}{k_0}\right)^2 \right]. 
\label{eq:couple2}
\end{equation}
Thus $\gamma(k) > 0$ (growth) for $ k < k_0$, $\gamma(k)$ has its
maximum value at $k=k_0$, and $\gamma(k)$ becomes strongly negative (damping)
as $k$ becomes large. 

  Our numerical implementation of this CCM model employs doubly periodic 
boundary conditions with periodicity lengths $L_x$ in $x$ and $L_y$ in $y$. 
The nonlinear map operator is applied at 
points on a square grid, while the spatial coupling operator (\ref{eq:couple1})
employs fast Fourier transforming from $\vec{x}$ to $\vec{k}$ and back.

  That $\xi(\vec{x})=0$ is an attractor can be seen by introducing an initial
perturbation at wavenumber $\vec{k}, \delta \xi_0$exp$(i\vec{k}\cdot\vec{x})$. 
Linearization of the CCM model about $\xi_0(\vec{x}) = 0$ then shows that this 
perturbation evolves with
time to $\delta \xi_n$exp$(i\vec{k}\cdot\vec{x})$, where $\delta \xi_n = \delta
\xi_0[M'(0)f(k)]^n, M'(\xi) \equiv \partial M / \partial \xi$. For the 
parameters we choose $M'(0)f(k) < 1$ for all $k$ (in particular $M'(0)f(k_0)<1$).
Thus the homogeneous state $\xi(\vec{x})=0$ is an attractor for the system. 
We also find that, as we had anticipated, for other initial conditions there
is another attractor which is spatio-temporally chaotic.

  Figure \ref{fig:stchaos1} shows the properties of the spatio-temporal chaos
produced by our model. Figure \ref{fig:stchaos1}(a) shows the spatial pattern
$\xi_n (\vec{x})$ at a representative time. 
This picture applies to a time $n = 45$ evolved from an initial condition where
$\xi_0 (\vec{x})$ was chosen randomly with uniform distribution between 
$\xi_0 = 0$ and $\xi_0 = 7.5$.
Visually, we observe that the 
pattern appears to have a characteristic scale of the order of $k_{0}^{-1}$.
This is confirmed by the wavenumber power spectrum, 
Fig. \ref{fig:stchaos1}(b). We note that the only length scales in our model 
are $k_{0}^{-1}$, the system size $L \sim L_x \sim L_y$, and the grid size 
$\delta$, and that, 
by our choice $ L \gg k_{0}^{-1} \gg \delta$, we had sought to obtain 
spatio-temporal chaos with properties independent of $L$ and $\delta$. 
Figure \ref{fig:stchaos1}(b), which evidences variation on the scale $k_0$, 
conforms with this expectation. Further discussion of the form observed for
$|\hat{\xi}(k)|^2$ is given in the appendix.
To characterize the temporal variation of the
patterns, Fig. \ref{fig:stchaos1}(c) shows a plot of the time correlation 
function $C(\tau)$ defined as
\begin{eqnarray}
C(\tau)&=&\frac{1}{N_x N_y} \sum_{i,j}^{N_x N_y} C_{i,j}(\tau), \\
C_{i,j}(\tau) &=& 
\frac{<(\xi_{i,j}(t+\tau)-\bar{\xi}_{i,j})(\xi_{i,j}(t)-\bar{\xi}_{i,j})>}
{<\xi_{i,j}^2(t) - \bar{\xi}_{i,j}^2>},
\label{eq:tcor}
\end{eqnarray}
where $<...>$ means time average, $\bar{\xi} = <\xi>, N_{x,y} = L_{x,y} / 
\delta$, and $(i,j)$ denotes the $(x,y)$ location of a grid point. 
As can be seen from Fig. \ref{fig:stchaos1}(c) the time correlation function 
decays to zero with increasing time $\tau$ (where $\tau$ is an integer), 
confirming that the temporal behavior is chaotic.

  We have also examined other parameter values for which 
(\ref{eq:map1})-(\ref{eq:couple2}) yields spatio-temporal chaos, and we find
behavior similar to that in Figs. \ref{fig:stchaos1}.

\section{PROPAGATION OF A FLAT FRONT}

\begin{figure}[b!]

\includegraphics[width= 7 cm]{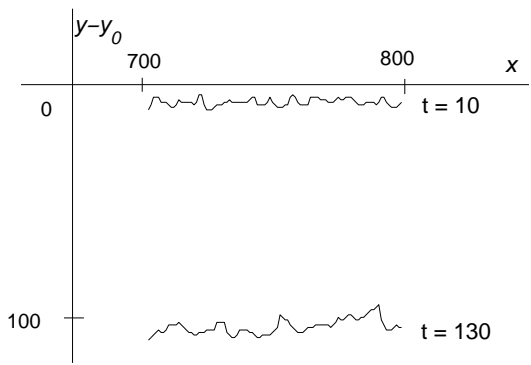}
\caption{Coarsening of a flat front. $L_x = 1024$.}
\label{fig:flat1}
\end{figure}

  The main objective of our investigation is to characterize the dynamics of
the interface between a spatio-temporally chaotic state and a stable steady
state. In our first set of simulations we focus on an initially flat interface,
$y = y_0 $ at $t = 0$.
After we generate an initial spatio-temporally  chaotic state 
(Fig. \ref{fig:stchaos1}(a)), 
we create the interface by setting the amplitude $\xi$ of all grid points with
$y < y_0$ to the stable steady state value $\xi = 0$.
During further iterations, the front between the chaotic and
steady state moves downward, i.e., the chaotic state propagates into the stable
steady state, and the front coarsens (see Fig. \ref{fig:flat1}). 
In order to examine the front dynamics for long times
and save computational time, we use a shifting method: On every iterate, we
reset $\xi$ to zero in the region adjacent to the bottom ($y=0$) of the 
periodic box, and, when the front comes close to $y = 0$, we shift the whole 
system upward in the \it y \rm direction.
Due to the periodic boundary conditions, after this shift there will be
a region below the front and above $y=0$ that is in the spatio-temporally
chaotic regime, and we then set $\xi=0$ in this region.

  After an initial transient, the scaling properties of the front are 
studied using the following definition of the interface width.
First, we calculated the average value of $\xi$ at fixed $ y $, $\tilde{\xi}
(y) = \int_0^{L_x} \xi (x,y) dx/L_x $. We then note that the basin boundary 
between the two attractors of the one dimensional map $M$ is at
the unstable fixed point, $ u = 0.4824 $ (Fig. \ref{fig:map}), 
and that the average $\xi$ for the spatio-temporally chaotic state (Fig. 
\ref{fig:stchaos1}(a)) is approximately four times the critical value. 
Thus, we define a lower boundary of the front, $ y_1 $, by 
$ \tilde{\xi}(y_1) = u $ and an upper boundary of the front, $ y_2 $, by 
$ \tilde{\xi}(y_2) = 3 u $. The width of the front is then defined as 
$(y_2 - y_1)$.

\begin{figure}[b!]

\includegraphics[width= 8 cm]{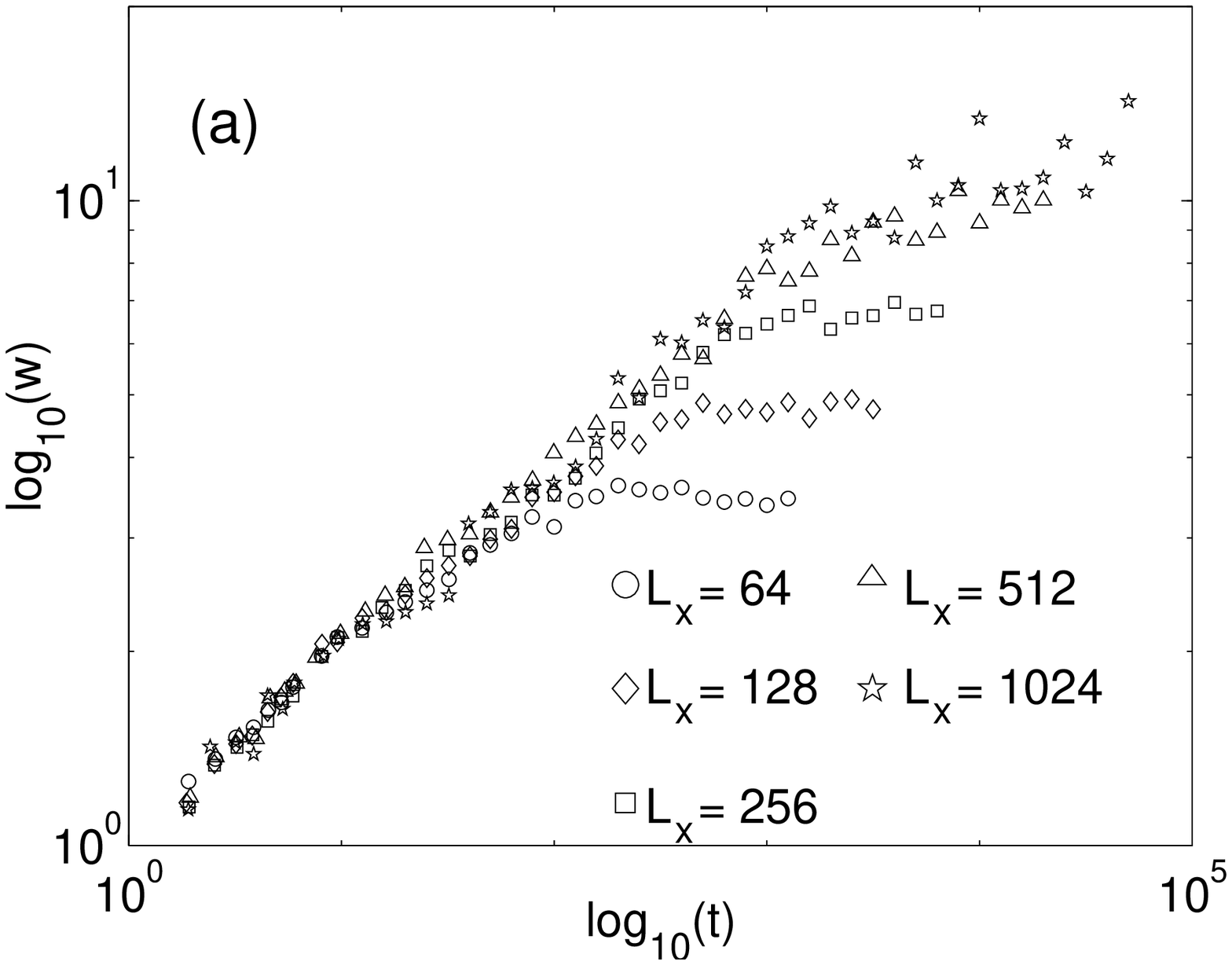}
\vspace{0.2cm}
\includegraphics[width= 8 cm]{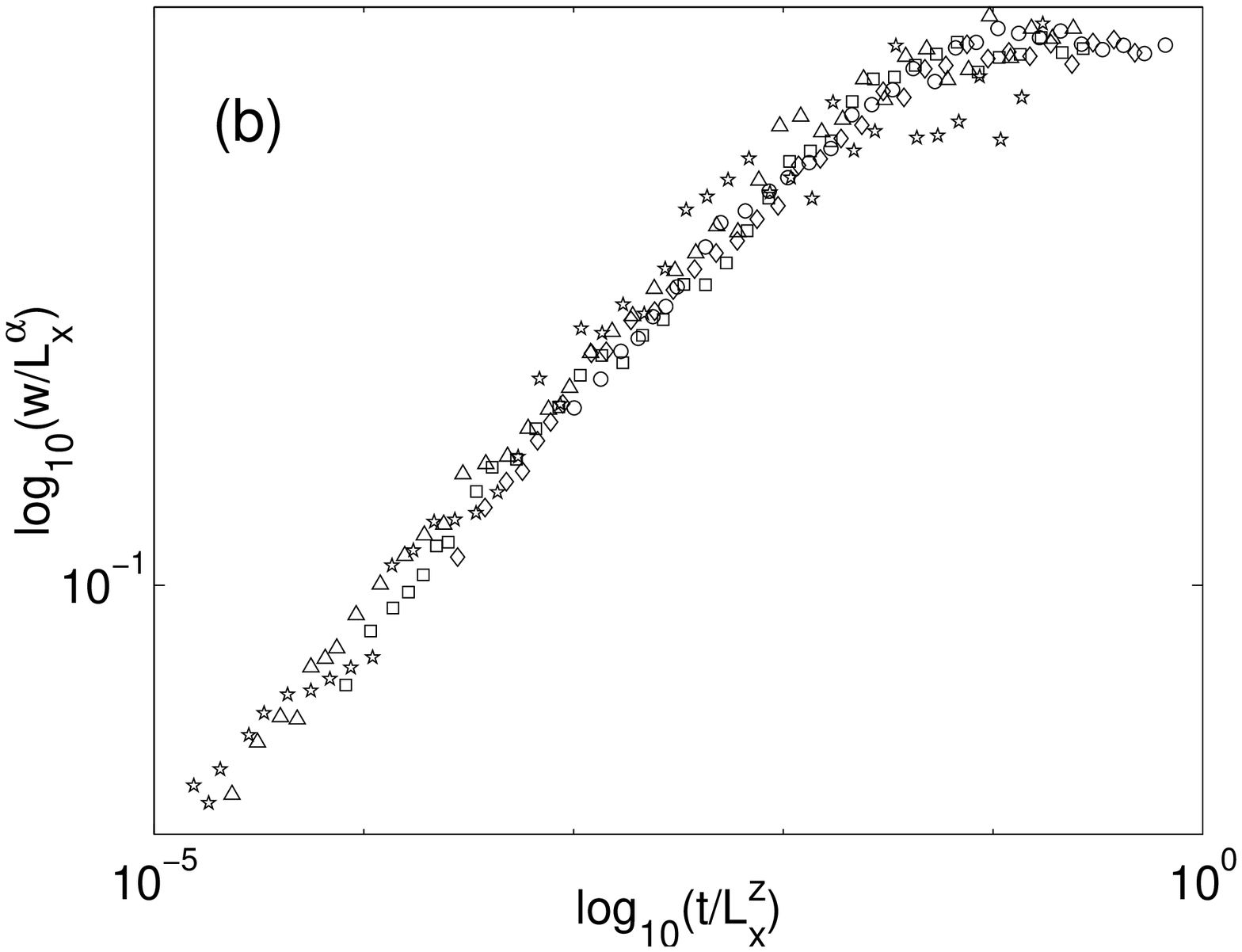}
\vspace{0.2cm}
\caption{Scaling of width with the size: (a) unscaled,
(b) scaled with $ \alpha = 0.49, z = 1.81$ from which $\beta =\alpha / z = 
0.27 $. Circles ($L_x = 64$), diamonds ($L_x = 128$), squares ($L_x = 256$), 
triangles ($L_x = 512$), and stars ($L_x = 1024 $). $L_y = 256$ for all cases.}
\label{fig:kpz}
\end{figure}

   Because of the inherent `noise' generated in the spatio-temporally chaotic
region, the proper quantity to study is the ensemble averaged mean of the front 
widths. We calculate ensembles using many different random initial conditions.
Our results for the ensemble averaged width $w$ of the front are obtained by 
averaging ($y_2 - y_1$) over 10 runs for the largest 
system $ L_x = 1024$ and over 500 runs for the smallest $ L_x = 64 $. 
The typical coarsening of the front is shown in Fig. \ref{fig:flat1}.   

  One observation from our simulations is that the propagation 
velocity of fronts is constant, except for a few transient initial iterations. 
That is, the velocity does not depend on time or system size. 
A constant propagation front velocity is also observed in the experiments in
Refs. \cite{Losert1} and \cite{Melnikov1}. 
To minimize the effect of the initial transient, 
we redefine time as the total increase in the area of the chaotic state. 

   As is typical for a front coarsening problem \cite{Barabasi1}, the time
and system size dependence of the mean front width $ w $ can be described by a 
scaling function $g(u)$,
\begin{equation}
w(t) = t^{\beta}g(\frac{t}{L_x^z}). 
\label{eq:sfunction}
\end{equation}
Here $ g(u) $ is constant for $ u \ll 1 $ and $ g(u) \approx u^{-\beta} $ for 
$ u \gg 1 $. For $ t \gg L_x^z $, the width saturates at $ w \sim L_x^{\alpha}$
where $ \alpha = \beta z$ is the roughness exponent.  
Barab\'{a}si and Stanley \cite{Barabasi1} have summarized the values of the 
scaling exponents $z$, $\alpha$, and $\beta$ that are obtained for several 
experimental systems as well as relevant theoretical results. 

  In Fig. \ref{fig:kpz}(a), we show $w$ versus $t$ for different system sizes.
These data show two characteristic regimes: power law growth, followed by 
saturation. The growth exponent, $ \beta $, is calculated by measuring
the slopes of straight line fits to the data before saturation, and the 
roughness exponents $ \alpha $ is calculated by comparing the saturation widths.
Figure \ref{fig:kpz}(b) shows $w/L_x^\alpha$ versus $t/L_x^z$, where $\alpha$
and $z$ have been adjusted to $\alpha = 0.49$ and $z = 1.81$ (corresponding to
$\beta = \alpha / z = 0.27$). Consistent with Eq. (\ref{eq:sfunction}), 
we find collapse of the data in Fig. \ref{fig:kpz}(a) to a single scaling 
function. The exponent values we obtain are roughly consistent with those of 
both the two dimensional Eden model ($\alpha \cong 0.5, \beta \cong 0.3$) and 
the two dimensional Kardar-Parisi-Zhang equation ($\alpha = 1/2, \beta = 1/3$)
\cite{Barabasi1,Kardar1}.
\begin{figure}[t]

\includegraphics[width= 7 cm]{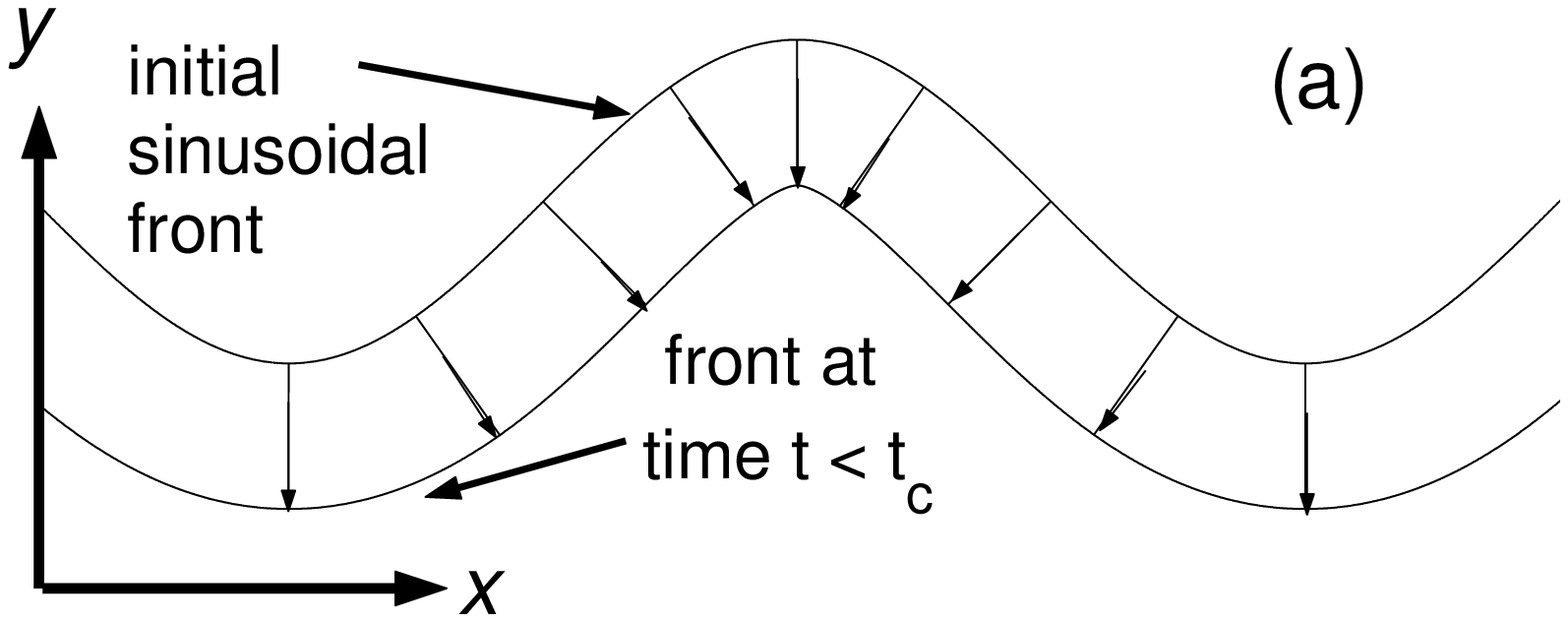}
\includegraphics[width= 7 cm]{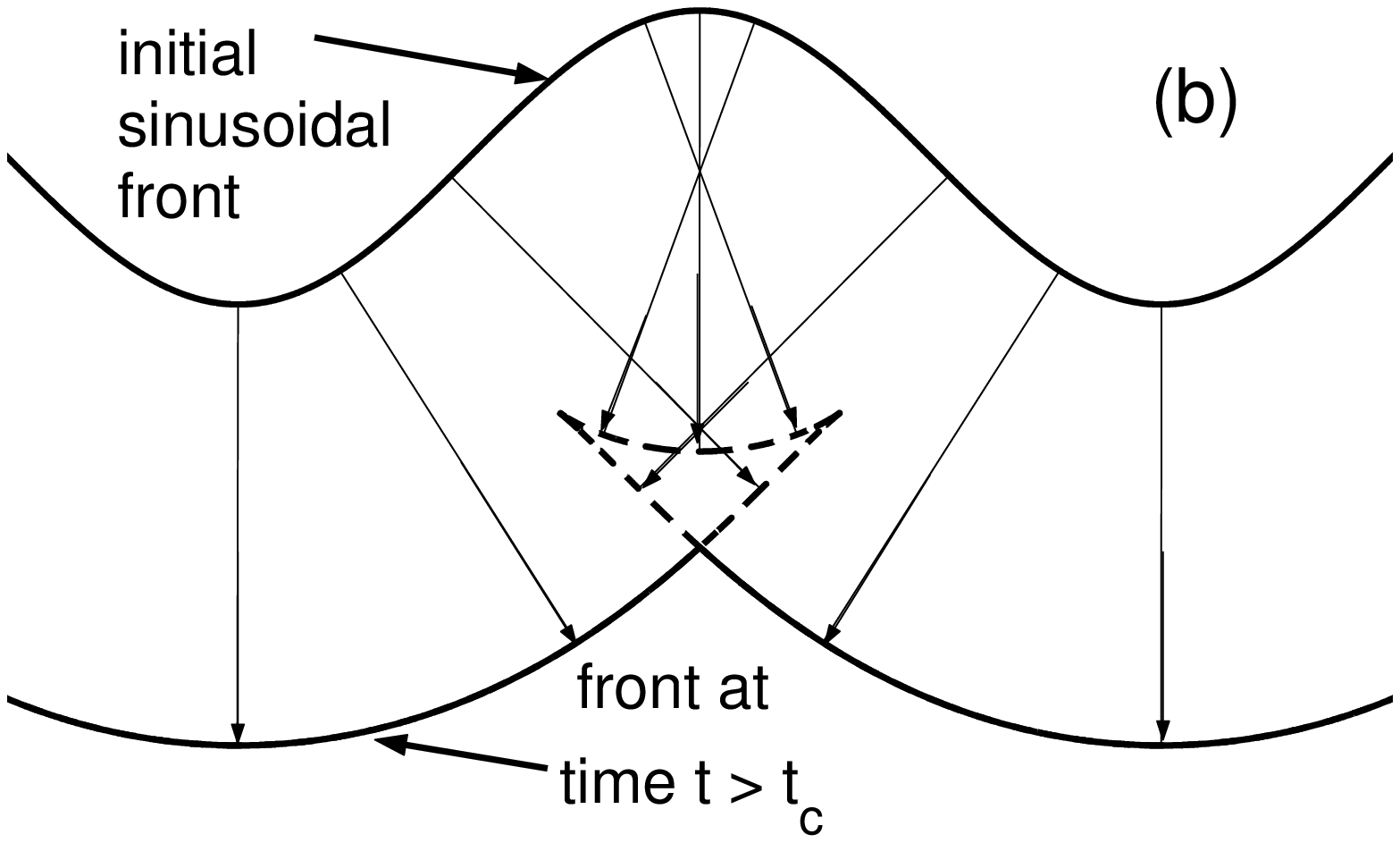}
\caption{Geometrical picture of the trajectories given by Eqs. (\ref{eq:cusp1})
and (\ref{eq:cusp2}) (a) for $t < t_c$ and (b) for $t > t_c$.}
\label{fig:cusp1}
\end{figure}

\section{EVOLUTION OF A NON-FLAT FRONT}
  We now consider the evolution of a front on large length scale. 
Specifically, we are
interested in the case where the front is initially not flat;
that is, the position of the front is initially given by
$y_0 = h(x_0)$. Furthermore, we assume that, as the front evolves, the scale
on which the front position varies, $l \approx (h/h')$, remains large compared 
to the front width,
\begin{equation}
l \gg w.
\label{eq:nonflat}
\end{equation}

  To analyze this situation, we consider that a point on the front moves with 
a normal velocity $\vec{v}$ whose magnitude, $|\vec{v}| = v$, is constant in
time. This assumption also implies that the direction of $\vec{v}$ following the
trajectory of a point on the front is constant in time. 
This is because the slope of the front, $dx / dy$, following a trajectory does
not change with time; i.e., it depends only on the initial location $x_0$ on
the front and not on $t$. This is illustrated by the construction in Fig. 
\ref{fig:cusp1}(a). As shown in the figure, the trajectory line segments are 
straight, are all of the same length, $vt$, and are normal to both the initial
front and to the evolved front.
Considering an initial front position given by $y_0 = h(x_0)$,
propagation at a velocity $\vec{v}$ normal to the front then yields
\begin{eqnarray}
\label{eq:cusp1}
x(t) & = & x_0 - (vt) h'(x_0)[1+(h'(x_0))^2]^{-1/2}, \\
\label{eq:cusp2}
y(t) & = & h(x_0) - vt [1+(h'(x_0))^2]^{-1/2}. 
\end{eqnarray}
At any given time Eqs. (\ref{eq:cusp1}) and (\ref{eq:cusp2}) specify the front 
position parametrically with $x_0$ as a parameter. As an example consider the
case of an initial sinusoidal undulation of the front $y_0 = h(x_0) =
C \cos(kx_0)$. As the front propagates, the initial sinusoidal curve becomes
distorted so that the maxima become sharp and the minima become broad. 
As can be seen from Fig. \ref{fig:cusp1} (a) this arises because of the 
converging 
(diverging) of trajectories that originate near maxima (minima).

  Past a critical time $ t = t_c $ a cusp develops at the maxima of the 
evolved front \cite{refcusp}. 
The cusp formation time is determined by noting that $ dx / dx_0 $ first 
becomes zero at $ t = t_c $. From Eq. (\ref{eq:cusp1}) we obtain
\begin{equation}
t_c = \frac{1}{k^2 C v}.
\label{eq:cusp3}
\end{equation}
For $ t > t_c $ there are pairs of values of $x_0$ for which the trajectories
given by (\ref{eq:cusp1}) and (\ref{eq:cusp2}) pass through each other. For a 
given time $t$ greater than $t_c$ we refer to the range of $x_0$ for which
this occurs as the unphysical range.
The development of the unphysical range is illustrated in Fig. \ref{fig:cusp1}
(b) where the dashed portion of the curve shows the result of plotting
(\ref{eq:cusp1}) and (\ref{eq:cusp2}) for the unphysical range.
 
\begin{figure}[t]
\includegraphics[width= 9 cm]{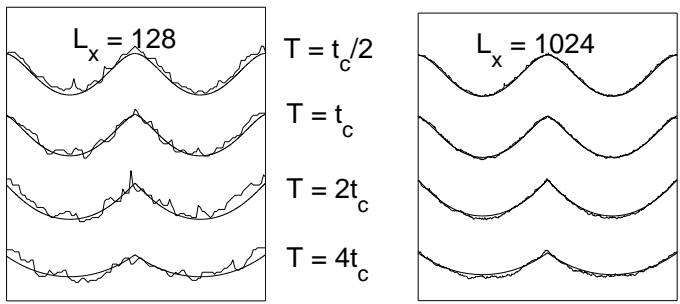}
\caption{Comparison between theoretical curves and our 2D simulations for 
two different system sizes. The front curve from our model is defined as the 
smallest $y$ value (for given $x$) at which $\xi = u$, where $u = 0.4824$ is
the basin boundary point depicted in Fig. \ref{fig:map}.}
\label{fig:cusp2}
\end{figure}

  Figure \ref{fig:cusp2} shows how cusps develop in time. 
In Fig. \ref{fig:cusp2}
the noisy front curves are from our simulations and the smooth curves are
from Eqs. (\ref{eq:cusp1}) and (\ref{eq:cusp2}) with $x_0$ restricted to the 
physical range. 
The two cases shown in Fig. \ref{fig:cusp2} (namely, $L_x = L_y = 128$ and
$L_x = L_y = 1024$) illustrate how front roughening becomes of less influence 
as (\ref{eq:nonflat}) becomes better satisfied. For both cases in Fig. 
\ref{fig:cusp2} the initial sinusoid has amplitude $kC = 1$ and wavenumber
$k = 4 \pi / L_x$. Thus from (\ref{eq:cusp3}) we have $t_c \sim 1/k \sim
L_x$. At times $t \sim t_c $, the roughening is a small effect if
$kw(t_c)$ is small. Since $ w \sim t^{\beta}$ (assuming $t/L_x^z \lesssim 1$),
we see that $kw(t_c) \sim
1/L_x^{1-\beta}$, and roughening will be inconsequential for the large scale 
front evolution if $L_x$ is sufficiently large. The good agreement of the 
$L_x =1024$ numerical results from our CCM model with the theory, Eqs.
(\ref{eq:cusp1}) and (\ref{eq:cusp2}), confirms that the front does indeed
propagate at constant velocity in a direction perpendicular to the interface.
In particular, for long length scales $l \gg w$ examined in Fig. 
\ref{fig:cusp2} we see no evidence for curvature dependence of the front
velocity.

\begin{figure}[h!]
\includegraphics[width= 6 cm]{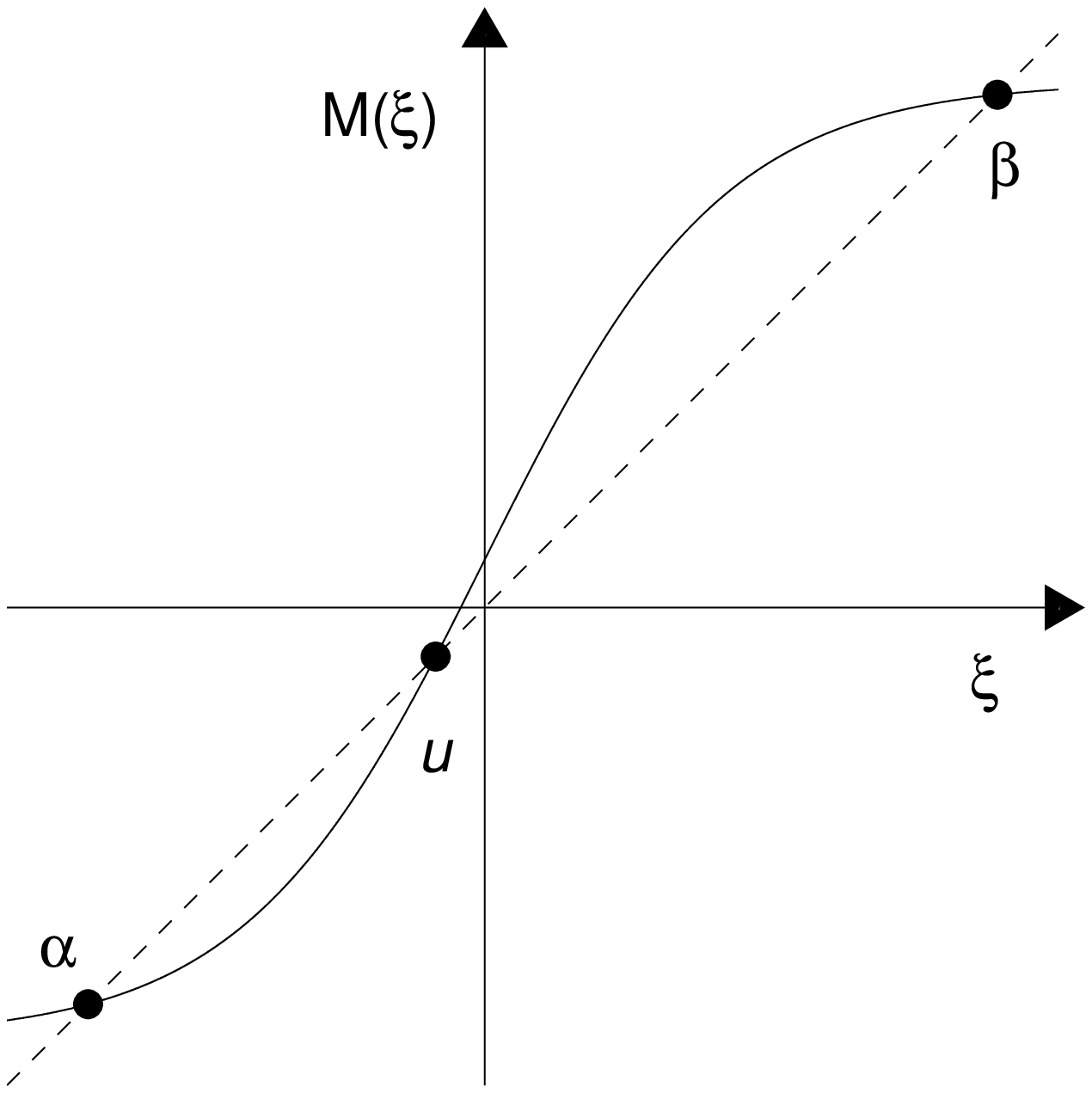}
\caption{The map Eq. (\ref{eq:newmap}).}
\label{fig:newmap}
\end{figure}

  As a comparison, we have also considered the propagation of a front between
two steady homogeneous states. In particular, replacing $M(\xi)$ in Eq.
\ref{eq:map2} by
\begin{equation}
M(\xi)=0.1 + \tanh (2 \xi),
\label{eq:newmap}
\end{equation}
we see (Fig. \ref{fig:newmap}) that there are two stable fixed points,
$\alpha$ and $\beta$, and one unstable fixed point $u$. Using 
(\ref{eq:newmap}), (\ref{eq:couple1}) and(\ref{eq:couple2}) with a sinusoidal
front with initialization of $\beta$ above the front and at $\alpha$ below the
front, we find that the front shape evolves smoothly (with no roughening)
according to Eqs. (\ref{eq:cusp1}) and (\ref{eq:cusp2}). Thus the cases of
a chaotic invading region and a nonchaotic invading region become similar
for large $L_x$ (e.g., $L_x \gg w \gg 1/k_0$).

  In conclusion, we have introduced a continuum coupled map model for the
study of the dynamics of a front separating a region of spatio-temporal chaos
from a stable steady region. This model is applicable to periodically forced
systems. We find that the front roughens and that this coarsening obeys a
scaling hypothesis, Eq. (\ref{eq:sfunction}). We also investigate the large
length scale evolution of a nonplanar front. We find that this evolution is
consistent with the hypothesis that, on large scale, the front velocity is 
constant and normal to the front. This hypothesis and our numerical simulations
indicate the formation of cusp structures in the front.

  This work was supported by ONR (Physics). The work of SCV was supported
by a research fellowship from the Alfred P. Sloan Jr. Foundation and by the
NSF (DMR9975533).

\section*{APPENDIX}
\begin{figure}[t!]
\includegraphics[width= 6 cm]{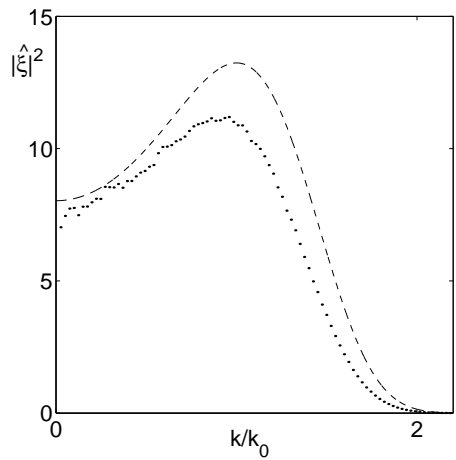}
\caption{$|\hat{\xi}|^2$ versus $k$ compared with Eq. (A.1).}
\label{fig:appendix1}
\end{figure}

  We now comment on the specific form that we have found for $|\hat{\xi}|^2$
(Fig. \ref{fig:stchaos1}(b)). In this connection we note that in the limit
of a wildly varying map $M(\xi)$ with  Lyapunov exponent approaching infinity
$\xi'_n(\vec{x})$ will be wildly varying in space. This is because small
variation of $\xi_n(\vec{x})$ with $\vec{x}$ are greatly amplified when $M$
is applied. Thus, in this limit, the spatial correlation function for 
$\xi'_n(\vec{x})$ will be a delta function, and 
$|\hat{\xi'}_n(k)|^2 = <(\xi'_n(\vec{x}))^2>$ 
independent of $\vec{k}$. Thus from (\ref{eq:couple1})
\[
~~~~~~~~~~~~~~~|\hat{\xi}(k)|^2 = <(\xi'_n(\vec{x}))^2>f^2(k),~~~~~~~~~~~~~(A.1)
\label{eq:appendix1}
\]
which is plotted in Fig. \ref{fig:appendix1} as the dashed line along with
the data from Fig. \ref{fig:stchaos1}(b). It is seen that Eq. (A.1) provides
a crude indication of the general form of $|\hat{\xi}(k)|^2$.

\begin{references}
\bibitem{Losert1}  W. Losert, D. G. W. Cooper, and J. P. Gollub, Phys. Rev. E
\bf 59\rm, 5855 (1999).
\bibitem{Melnikov1} I. V. Melnikov, D. A. Egolf, S. Jeanjean, B. B. Plapp,
and E. Bodenschatz, to appear in ``Stochastic Dynamics and Pattern Formation
in Biological and Complex Systems", edited by S. Kim, K. J. Lee, T. K. Lim,
and W. Sung, AIP Conference Proceeding 1999.
\bibitem{Shankar1}  S. C. Venkataramani, and E. Ott, Phys. Rev. Lett.
 \bf 80\rm, 3495 (1998).
\bibitem{Barabasi1}  A.-L. Barab\'{a}si and H. E. Stanley, \it Fractal Concepts
in Surface Growth \rm (Cambridge University Press, Cambridge, 1995). 
\bibitem{Description1} This general viewpoint was also adopted in Ref.
\cite{Shankar1} where, motivated by experimentally observed phenomena 
[F. Melo, P. B. Umbanhowar, and H. L. Swinney, Phys. Rev. Lett. \bf 75 \rm, 
3838 (1995)], other choices for $M$ and $f$ were employed. In that case the
objective was to check the hypothesis that period doubling in conjunction with
pattern formation at a preferred scale were the essential ingredients 
necessary to explain the observed bifurcations and the evolution of time 
periodic patterns occurring in a vertically oscillated granular layer
of the order of 10 grains thick.
\bibitem{Kardar1}  M. Kardar, G. Parisi, and Y. C. Zhang, Phys. Rev. Lett. 
\bf 56\rm, 889 (1986).
\bibitem{refcusp} Cusp formation also occurs for the Kardar-Parisi-Zhang
equation; see Fig. 1 of Ref. \cite{Kardar1}. 
\end{references}
\end{document}